\documentclass[twocolumn,english,prb]{revtex4-2}
\usepackage{color}
\usepackage[utf8]{inputenc}
\usepackage{graphicx}
\usepackage[T1]{fontenc}
\usepackage{float}
\usepackage{textcomp}
\usepackage{amsmath}
\usepackage{amssymb}
\usepackage{graphicx}
\usepackage{esint}
\usepackage{xcolor}
\usepackage{multirow}
\usepackage{dsfont}
\usepackage{physics}
\usepackage{bbold}

\usepackage[colorlinks=true,hyperfootnotes=true,breaklinks=true]{hyperref}
\usepackage[capitalise]{cleveref}
\begin{document}
\title{Spin SWAP operation in double quantum dots at the LaAlO$_3$/SrTiO$_3$ interface.}

\author{A. Sierant}
\affiliation{AGH University of Krak\'ow,
Faculty of Physics and Applied Computer Science,
al. Mickiewicza 30, 30-059 Krak\'ow, Poland}

\author{J. Czarnecki}
\affiliation{AGH University of Krak\'ow,
Faculty of Physics and Applied Computer Science,
al. Mickiewicza 30, 30-059 Krak\'ow, Poland}

\author{B. Szafran}
\affiliation{AGH University of Krak\'ow,
Faculty of Physics and Applied Computer Science,
al. Mickiewicza 30, 30-059 Krak\'ow, Poland}

\author{P. Wójcik}
\email{pawelwojcik@agh.edu.pl}
\affiliation{AGH University of Krak\'ow,
Faculty of Physics and Applied Computer Science,
al. Mickiewicza 30, 30-059 Krak\'ow, Poland}

\begin{abstract}
Progress in the fabrication of nanoscale transition-metal-oxide heterostructures makes these platforms promising candidates for the realization of spin qubits, mainly due to the \(d\)-character of their electronic structures, which could potentially result in a reduction of hyperfine interactions and spin decoherence. Here, we present a systematic study of spin control within the SWAP operation in double quantum dots embedded in a two-dimensional electron gas at the LaAlO$_3$/SrTiO$_3$ interface. Our analysis starts with a study of single-electron spin dynamics, focusing on the influence of spin-orbit and interorbital coupling on the spin evolution. In this case, our findings are supported by semiclassical calculations based on the Bloch equations, which show good agreement with full quantum mechanical simulations. We then simulate the SWAP operation by analyzing the crossover between two regimes: (i) large quantum dots, where the electronic structure is dominated by the \(d_{xy}\) orbitals and the spin dynamics is affected primarily by Rashba-type spin-orbit interaction; and (ii) small quantum dots, where higher-energy orbitals $d_{xz/yz}$ contribute to the electronic structure, leading to a significant reduction in the SWAP fidelity. In the first regime, particularly relevant from the application point of view, we analyze in detail the anisotropy of the SWAP operation induced by the spin-orbit coupling.
\end{abstract}

\date{\today}
\maketitle
\section{Introduction}

For several decades, quantum dots (QDs)~\cite{Kouwenhoven2002,burkard1999coupled} have attracted considerable attention due to their potential as fundamental building blocks for quantum information technologies~\cite{Burkard2023,loss1998quantum}. In particular, recent studies have focused on constructing effective spin qubits, where the precise control of electron spin confined in QD~\cite{Petta2005,Koppens2005} is strongly influenced by the host material, as it determines the strengths of the spin-orbit (SO) coupling~\cite{Rashba,Rashba2003,Bychkov1984} and the hyperfine interaction~\cite{Hanson2007}. The latter constitutes a primary source of spin decoherence~\cite{Loss2002,Hanson2007,Koppens2008}, limiting the suitability of a given material platform for use in quantum computing technologies~\cite{loss1998quantum}.

To date, the manipulation of electron spin has been successfully demonstrated in semiconductor-based QDs~\cite{Petta2005, Nowak2012,Nadj-Perge2010, Nadj-Perge2012_2,Koppens2005,Koppens2006}, which allows for electrical control of spin state~\cite{huang2024high,loss1998quantum, pla2013high} and offers a promising pathway toward scalable quantum devices. However, semiconductor QD technologies are fundamentally limited by spin decoherence, which is particularly pronounced for conduction-band electrons. For this reason, alternative platforms are being intensively explored, including $p$-type semiconductors~\cite{Fischer2008,Maurand2016}, graphene~\cite{Trauzettel2007}, and transition-metal dichalcogenides~\cite{Li2017}. 
Among them, one particularly promising and still unexplored material platform is provided by transition-metal oxide interfaces~\cite{Chen2024}, which are characterized by the electronic structure of $d$-electrons. In such systems, the hyperfine interaction with the nuclear spin bath is strongly suppressed, since its dominant contribution - the contact hyperfine term - is proportional to the squared amplitude of the electronic wave function at the nuclear site~\cite{Loss2002}. For $d$ orbitals, this amplitude vanishes, leading to a substantial reduction of this decoherence channel.

Within the family of transition-metal oxides, the LaAlO$_3$/SrTiO$_3$ (LAO/STO) interface stands out as one of the most promising and extensively explored heterostructures~\cite{Chen2024}. The two-dimensional electron gas (2DEG) at the LAO/STO interface  exhibits a unique combination of characteristics, including high carrier mobility~\cite{Ohtomo2004}, strong SO coupling~\cite{Diez2015,Rout2017,Caviglia2010,Shalom2010,Yin2020,Singh2017,Hurand2015}, the gate-tunable superconductivity~\cite{Reyren2007, Joshua2012, Maniv2015,
Biscaras2012, Monteiro2019, Monteiro2017, wojcik2021impact} and magnetic ordering~\cite{Li2011,Bert2011}.
The first experimental realization of electrostatically defined LAO/STO QDs has already been reported in Ref.~\cite{Jespersen2020}, demonstrating pronounced Coulomb diamonds~\cite{Jespersen2020}. Remarkably, some of these experiments provide evidence of the attractive interaction between electrons, which could shed light on the pairing mechanism in the superconducting LAO/STO interface~\cite{Guenevere2017}.

Although the experimental demonstration of electrical spin control in LAO/STO QDs has not yet been achieved, theoretical studies indicate the possibility of spin manipulation via electric-dipole spin resonance (EDSR)~\cite{szafran2024electric,Szafran_manipulation_2024}. The next step toward realizing universal quantum gates involves implementing controllable spin exchange between electrons confined in coupled QDs. Note, however, that the SWAP operation in LAO/STO double QDs raises important questions regarding the role of the multiorbital $d$ character of the LAO/STO 2DEG and the impact of SO coupling, which can introduce anisotropic corrections to the SWAP operation~\cite{Nowak2010}.

In this work, we systematically study the spin dynamics during the SWAP operation in double QDs formed at the LAO/STO interface, focusing on the crossover between two distinct regimes: (i) large QDs, where the electronic structure is mainly determined by the \(d_{xy}\) orbitals and the spin dynamics are predominantly governed by Rashba-type SO coupling, and (ii) small QDs, where higher-energy orbitals become relevant,  producing pronounced distortions of the SWAP operation. Within the experimentally accessible regime, we further investigate in detail the anisotropy of the SWAP operation induced by the SO coupling.

The manuscript is organized as follows. In Sec.~\ref{sec:theory}, we introduce the real-space tight-binding model used to describe the 2DEG at the (001) LAO/STO interface and outline the configuration interaction (CI) method employed for two-electron calculations. Sec.~\ref{sec:single_electron} presents an analysis of the single-electron spin dynamics in double QDs. In Sec.~\ref{sec:swap}, we investigate the SWAP operation. The anisotropy of SWAP induced by the SO coupling is discussed in Sec.~\ref{sec:swap_aniso}, while Sec.~\ref{sec:swap_scale} demonstrates the applicability of the scaled model for simulation of the spin dynamics. Finally, a summary is provided in Sec.~\ref{sec:summary}.

\section{Theoretical model}
\label{sec:theory}
\subsection{The single electron Hamiltonian for the LAO/STO interface}
We consider a 2DEG created at the $(001)$ LAO/STO  interface, where the conduction band is described by $t_{2g}$ orbitals of Ti ($d_{xy}$, $d_{zy}$, $d_{xz}$). The system is characterized by the $\mathbf{L} \cdot \mathbf{S}$ atomic SO coupling as well as the Rashba SO interaction occurring as a result of the mirror symmetry breaking at the interface. To describe the LAO/STO 2DEG, we use a tight-binding Hamiltonian, which in the wave vector space is given by~\cite{Diez2015}
\begin{equation}
    \hat{H}_{\mathbf{k}}=\sum _{\mathbf{k}} \hat{C}^\dagger_{\mathbf{k}} ( \hat{H}_{0}+\hat{H}_{RSO}+\hat{H}_{SO}+\hat{H}_B ) \hat{C}_{\mathbf{k}},
    \label{eq:Hamiltonian_k_space}
\end{equation}
where $\hat{C}_{\mathbf{k}}=(\hat{c}_{\mathbf{k},xy}^{\uparrow}, \hat{c}_{\mathbf{k},xy}^{\downarrow}, \hat{c}_{\mathbf{k},xz}^{\uparrow}, \hat{c}_{\mathbf{k},xz}^{\downarrow}, \hat{c}_{\mathbf{k},yz}^{\uparrow}, \hat{c}_{\mathbf{k},yz}^{\downarrow})^{T}$ is the vector of annihilation operators for electrons with spin $\sigma=\uparrow,\downarrow$ on the orbital $d_{xy},d_{xz},d_{yz}$, in the state $\mathbf{k}$.  

In Eq.~(\ref{eq:Hamiltonian_k_space}), $\hat{H}_0$ corresponds to the kinetic energy
and takes the form
\begin{equation}
\hat{H}_{0}=
\left(
\begin{array}{ccc}
 \epsilon^{xy}_{\mathbf{k}} & 0 & 0\\
 0 & \epsilon^{xz}_{\mathbf{k}} &  \epsilon^h_{\mathbf{k}} \\
 0 & \epsilon^h_{\mathbf{k}}  &  \epsilon^{yz}_{\mathbf{k}}
\end{array} \right) \otimes \hat {\sigma} _0\;,
\end{equation}
with  
\begin{equation}
\begin{split}
    \epsilon^{xy}_{\mathbf{k}}&=4t_l-2t_l\cos{k_x}-2t_l\cos{k_y}-\Delta_E,\\
    \epsilon^{xz}_{\mathbf{k}}&=2t_l+2t_h-2t_l\cos{k_x}-2t_h\cos{k_y},\\
    \epsilon^{yz}_{\mathbf{k}}&=2t_l+2t_h-2t_h\cos{k_x}-2t_l\cos{k_y},
\end{split}
\label{eq:H0}
\end{equation}
and the hybridization energy
\begin{equation}
\epsilon^h_{\mathbf{k}}=2t_d\sin{k_x}\sin{k_y}. 
\end{equation}

The Hamiltonian $\hat{H}_{SO}$ describes the atomic SO interaction, represented by the $\mathbf{L} \cdot \mathbf{S}$ coupling term, and can be written as~\cite{Khalsa2013}
\begin{equation}
\hat{H}_{SO}= \frac{\Delta_{SO}}{3}
\left(
\begin{array}{ccc}
0 & i \sigma _x & -i \sigma _y\\
-i \sigma _x & 0 & i \sigma _z \\
i \sigma _y & -i \sigma _z & 0
\end{array} \right) \;,
\label{eq:hso}
\end{equation}
where $\Delta _{SO}$ determines the strength of the atomic SO energy and $\sigma _{x},\sigma_{y},\sigma_{z}$ are the Pauli matrices.

As a consequence of mirror-symmetry breaking, the SO interaction acquires an additional Rashba-type contribution, denoted by \(\hat{H}_{\mathrm{RSO}}\), which for the LAO/STO interface takes the form
\begin{equation}
\hat{H}_{RSO}= \Delta_{RSO}
\left(
\begin{array}{ccc}
0 & i \sin{k_y} & i \sin{k_x}\\
-i \sin{k_y} & 0 & 0 \\
-i \sin{k_x} & 0 & 0
\end{array} \right) \otimes \hat {\sigma} _0\;,
\label{eq:rso}
\end{equation}
where $\Delta _{RSO}$ determines the energy of the Rashba SO coupling.

Finally, the coupling of the external magnetic field to the spin and orbital momentum of electrons is taken into account by the Hamiltonian
\begin{equation}
\hat{H}_B=\mu_B(\mathbf{L}\otimes \sigma_0+g\mathds{1}_{3\times 3} \otimes \mathbf{S})\cdot \mathbf{B}/\hbar,
\label{eq:Hb}
\end{equation}  
where $\mu_B$ is the Bohr magneton, $g$ is the Land\'e factor, $\mathbf{S}=\hbar \pmb{\sigma}/2$ with $\pmb{\sigma}=(\sigma_x,\sigma_y,\sigma_z)$ and $\mathbf{L}=(L_x,L_y,L_z)$ with
\begin{equation}
\begin{split}
 L_x&= \left ( 
 \begin{array}{ccc}
  0 & i & 0 \\
  -i & 0 & 0 \\
  0 & 0 & 0 
 \end{array}
 \right ), 
 L_y= \left ( 
 \begin{array}{ccc}
  0 & 0 & -i \\
  0 & 0 & 0 \\
  i & 0 & 0 
 \end{array}
 \right ), 
 L_z= \left ( 
 \begin{array}{ccc}
  0 & 0 & 0 \\
  0 & 0 & i \\
  0 & -i & 0 
 \end{array}
 \right ).
 \end{split}
\end{equation}

To determine the electronic spectrum of the QD, the Hamiltonian (\ref{eq:Hamiltonian_k_space}) is discretized  in real space on a square  lattice with mesh constant $a = 0.39$~nm, corresponding to the distance between Ti atoms at the LAO/STO interface
\begin{equation}
\begin{split}
\hat{H} = & \sum_{\mu, \nu} \hat{C}^{\dag}_{\mu, \nu} 
(\hat{H}^0 + \hat{H}_{SO} + \hat{H}_B + V_{\mu,\nu}\mathbf{1}_{3\times 3}\otimes \sigma_0) 
\hat{C}_{\mu, \nu}   \\
&+ \sum_{\mu, \nu} \hat{C}^{\dag}_{\mu +1, \nu} 
\hat{H}^x \hat{C}_{\mu, \nu} 
+ \sum_{\mu, \nu} \hat{C}^{\dag}_{\mu, \nu +1} 
\hat{H}^y \hat{C}_{\mu, \nu} \\
&+ \sum_{\mu, \nu} \hat{C}^{\dag}_{\mu +1, \nu -1} 
\hat{H}_{\text{mix}} \hat{C}_{\mu, \nu}
- \sum_{\mu, \nu} \hat{C}^{\dag}_{\mu +1, \nu +1} 
\hat{H}_{\text{mix}} \hat{C}_{\mu, \nu} + h.c.
\end{split}
\label{eq:Hamiltonian_real_space}
\end{equation}
where $\hat{C}_{\mu, \nu} = (\hat{c}^{\uparrow}_{\mu, \nu, xy},\hat{c}^{\downarrow}_{\mu, \nu, xy},\hat{c}^{\uparrow}_{\mu, \nu, xz},\hat{c}^{\downarrow}_{\mu, \nu, xz}, \hat{c}^{\uparrow}_{\mu, \nu, yz},\hat{c}^{\downarrow}_{\mu, \nu, yz})^T$ is the vector of electron annihilation operators for an electron with spin $\sigma=\uparrow, \downarrow$ on orbitals $d_{xy}, d_{xz}, d_{yz}$ at positions ($\mu, \nu$).

Note that Eq.~(\ref{eq:Hamiltonian_real_space}) is supplemented with respect to Eq.~(\ref{eq:Hamiltonian_k_space}) by an additional term, $V_{\mu,\nu}$, corresponding to the confinement potential that defines the QD. The remaining terms in Eq.~(\ref{eq:Hamiltonian_real_space}) result from the discretization and include the onsite energy
\begin{equation}
\hat{H}^0 =\;
\begin{pmatrix}
4t_l - \Delta_E & 0 & 0 \\
0 & 2t_l + 2t_h & 0 \\
0 & 0 & 2t_l + 2t_h
\end{pmatrix}
\otimes \hat{\sigma}_0;
\end{equation}
and the hopping energies related to the kinetic term and Rashba SO coupling~\cite{Perroni2019}
\begin{equation}
    \hat{H}^x
=
\begin{pmatrix}
 -t_l & 0 & 0 \\
 0 & -t_l & 0 \\
 0 & 0 & -t_h
\end{pmatrix}
\otimes \hat{\sigma}_0
+
\frac{\Delta_{\text{RSOC}}}{2}
\begin{pmatrix}
0 & 0 & -1 \\
0 & 0 & 0 \\
1 & 0 & 0
\end{pmatrix}
\otimes \hat{\sigma}_0,
\end{equation}

\begin{equation}
    \hat{H}^y
=
\begin{pmatrix}
 -t_l & 0 & 0 \\
 0 & -t_h & 0 \\
 0 & 0 & -t_l
\end{pmatrix}
\otimes \hat{\sigma}_0
+
\frac{\Delta_{\text{RSOC}}}{2}
\begin{pmatrix}
0 & -1 & 0 \\
1 & 0 & 0 \\
0 & 0 & 0
\end{pmatrix}
\otimes \hat{\sigma}_0.
\end{equation}

Finally, the hybridization between the orbitals $d_{xz}$ and $d_{yz}$ is included by
\begin{equation}
  \hat{H}_{mix} = \frac{\Delta_{t_d}}{2}
\begin{pmatrix}
0 & 0 & 0 \\
0 & 0 & 1 \\
0 & 1 & 0
\end{pmatrix}
\otimes \hat{\sigma}_0.
\end{equation}

In our calculations, we adopt the tight-binding parameters $t_l = 875\,$meV, $t_h = 40\,$meV, $t_d = 40\,$meV, and $\Delta_E = 47\,$meV, as reported in Ref.~\cite{Maniv2015}. We use a Landé $g$-factor of $g = 3$~\cite{Ruhman2014}, and spin–orbit coupling parameters $\Delta_{\mathrm{SO}} = 10\,$meV and $\Delta_{\mathrm{RSO}} = 20\,$meV, consistent with experimentally determined values~\cite{Caviglia2010,Yin2020}.

\subsection{Two-electron problem}
In the paper we consider the problem of two electrons in the double LAO/STO QD. The Hamiltonian of two electrons can be written as
\begin{equation}
    \hat{H}_2 = \hat{H}(1) + \hat{H}(2) + \frac{e^2}{4\pi\epsilon_0\epsilon r_{12}}
    \label{eq:two_electrons}
\end{equation}
where $\hat{H}$ is the single-electron Hamiltonian given by Eq.~(\ref{eq:Hamiltonian_real_space}). In Eq.~(\ref{eq:two_electrons}), the last term represents the Coulomb interaction, with the strength determined by the dielectric constant $\epsilon$. Although $\epsilon$ in STO is known to depend strongly on the electric field and temperature, reaching values as high as 20000, the large electric field present near the LAO/STO interface where the 2DEG is formed reduces its value to the range of $100$--$300$~\cite{szafran2024electrical}. In our calculations, we assume $\epsilon = 100$.

The eigenproblem of electron pair is solved using the exact diagonalization method in a basis of antisymmetrized products of single-electron eigenfunctions constructed from the Ti $3d$ orbitals
\begin{equation}
\begin{aligned}
\Psi_q(x,y,\sigma) 
&= \sum_{j} a_j^q d_j(x,y,\sigma) \\
&= \sum_{r_j,o_j,s_j} a_j^q d_{r_j,o_j}(x,y) S_{s_j}(\sigma).
\end{aligned}
\end{equation}
where the summation runs over the position of ions $r_j$, orbitals $o_j$ on the ion, and the $z$ component of the spin indexed by $s_j$ (S stands for the spin-up or spin-down eigenstate).

Determining the energy spectrum of and electron pair requires the evaluation of the Coulomb integrals
\begin{eqnarray}
&&I_{q_1q_2q_3q_4}=\langle \Psi_{q_1}(1)\Psi_{q_2}(2)|\frac{1}{r_{12}}|\Psi_{q_3}(1)\Psi_{q_4}(2)\rangle \\ \nonumber &=& 
\sum_{j_1,j_2,j_3,j_4}(a_{j_1}^{q_1}a_{j_2}^{q_2})^*a_{j_3}^{q_3}a_{j_4}^{q_4}\langle d_{j_1}(1) d_{j_2}(2)|\frac{1}{r_{12}}|d_{j_3}(1) d_{j_4}(2)\rangle,
\end{eqnarray}
where the integral over the spin-orbitals that appears in the sum is calculated based on the formula
\begin{eqnarray}
&&\langle d_{j_1}(1) d_{j_2}(2)|\frac{1}{r_{12}}|d_{j_3}(1) d_{j_4}(2)\rangle\nonumber = \\&&\delta(r_{j_1},r_{j_3})\delta(r_{j_2},r_{j_4}) \delta(s_{j_1},s_{j_3})\delta(s_{j_2},s_{j_4}) \times \nonumber \\&&
\bigg[ \left(1-\delta(r_{j_1},r_{j_2})\right)\frac{1}{|r_{j_1}-r_{j_2}|}\delta(o_{j_1},o_{j_3})\delta(o_{j_2},o_{j_4})
+ \nonumber \\&& \delta(r_{j_1},r_{j_2}) \varepsilon(o_{j_1},o_{j_2},o_{j_3},o_{j_4}) \bigg ],
\label{CI:integral}
\end{eqnarray}
where 
\begin{equation}
\varepsilon(o_{j_1},o_{j_2},o_{j_3},o_{j_4})=\langle d_{j_1}(1)d_{j_2}(2)|\frac{1}{r_{12}}|d_{j_3}(1)d_{j_4}(2) \rangle
\end{equation}
is a single-ion integral, evaluated using Monte-Carlo integration. Using hydrogen-like orbitals with an effective atomic number \(Z^* = 3.65\) for Ti and the normalization factor \(N\), the orbitals are defined as
$ d_{xy} = N\, e^{-Z^* r /3} xy \rightarrow 1$, $d_{xz} = N\, e^{-Z^* r /3} xz \rightarrow 2$, and 
$d_{yz} = N\, e^{-Z^* r /3} yz \rightarrow 3$. In atomic units, the non-zero on-site integrals are
$\varepsilon(i,i,i,i) = 0.336$, $\varepsilon(i,j,i,j) = 0.306$, $\varepsilon(i,j,j,i) = 0.015 \quad (i \neq j)$. The other single-ion integrals are zero. 

In the calculations we use up to 22 single electron states which form 231 Slater determinants as a basis for the two-electron problem.
\begin{figure}[!t]
    \includegraphics[width=\linewidth]{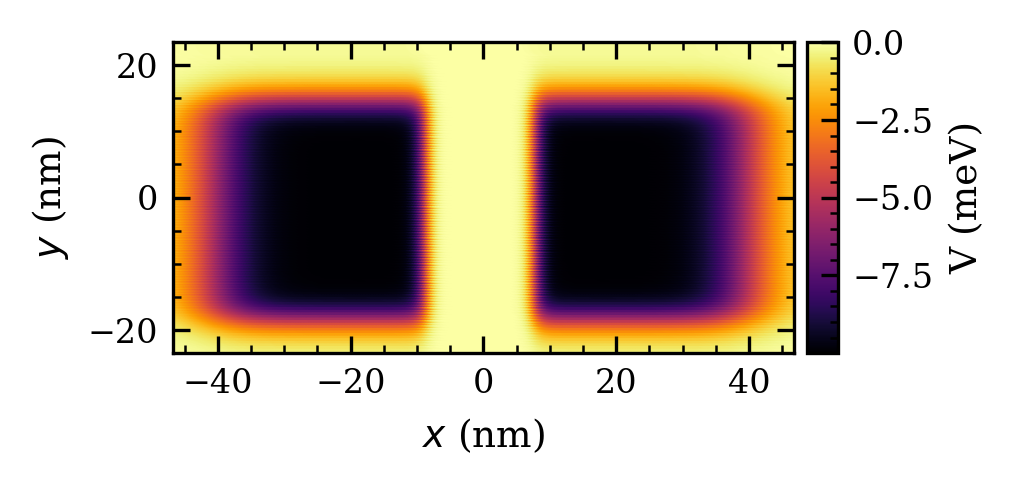}
    \caption{The confining potential in the double LAO/STO QD for the following set of parameters: 
\(\mu = 6\), \(V_0 = 10~\mathrm{meV}\), \(V_b = 10~\mathrm{meV}\), and \(R_b = \mathrm{43}a\).}
    \label{fig:potential}
\end{figure}

\section{Results}
In this section, we analyze the spin dynamics of a single electron and two electrons confined in the double LAO/STO QD. To describe two coupled dots, we employ a confining potential of the form
\begin{equation}
\begin{split}
    V(x,y) = &-V_0 / [(1+[x^2/(1.2 d)^2]^{\mu})(1+[y^2/(0.5 d)^2]^{\mu})] \\
    &+ V_b / [(1+[x^2/R_b^2]^{\mu})(1+[y^2/(0.5 d)^2]^{\mu})],
    \end{split}
\label{eq:potential}
\end{equation}
where \(\mu\) controls the steepness of the potential (\(\mu = 6\)), \(V_0 = 10~\mathrm{meV}\) is the depth of the potential well, \(V_b\) is the height of the barrier between the dots, \(R_b\) is the width of the barrier, and $d$ is the size of a single QD. The coefficients 1.2 and 0.5 appearing with  $d$ are included to preserve the symmetric shape of a single quantum dot and are related to the fact that position is measured relative to the point $(0,0)$. The schematic illustration of the potential profile for an exemplary set of parameters is shown in Fig.~\ref{fig:potential}.

\begin{figure}
    \centering
    \includegraphics[width=0.8\linewidth]{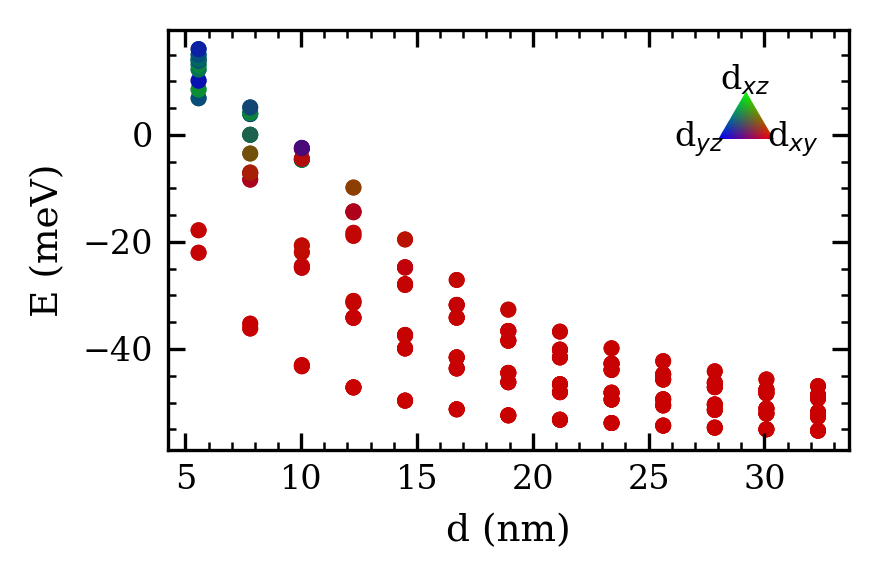}
    \caption{Single electron spectrum of double LAO/STO QD as a function of quantum dot size $d$.
    Contributions from different orbitals $d_{xy}$, $d_{xz}$ and $d_{yz}$ are marked using an RGB color scheme. Results for the confining potential parameters $V_b =70$ meV, $V_0=10$ meV and $R_b=70$ meV.}
    \label{fig:single_el_spectrum_as_size}
\end{figure}

\subsection{Single electron spin dynamics}
\label{sec:single_electron}
Let us start with the analysis of single-electron spin dynamics. As an initial condition, we consider an electron confined in the left QD, prepared as a superposition of the bonding and antibonding orbitals corresponding to the two lowest-energy eigenstates of the Hamiltonian (\ref{eq:Hamiltonian_real_space}). A weak magnetic field ($10^{-8}~\mathrm{T}$) is applied to polarize the initial spin state antiparallel to the $x$, $y$, or $z$ axis. Note that the field strength on the order of $10^{-8}$~T is sufficient to define the initial spin orientation while remaining negligible for the spin dynamics, ensuring that the time evolution of the spin is governed entirely by the SO interaction.

The solution of the eigenproblem for the Hamiltonian (\ref{eq:Hamiltonian_real_space}) is  used to construct the time evolution according to the Schr\"odinger equation,
$i\hbar \frac{d\Psi(t)}{dt} = H \Psi(t)$. The time-dependent wave function is given by
\begin{equation}
\label{eq:evolution}
\Psi(t) = \sum_m c_m \, e^{-i E_m t/\hbar} \Psi_m,
\end{equation}
where $\Psi_m$ and $E_m$ denote the eigenstates and eigenvalues, respectively, and the coefficients $c_m$ are determined from the initial condition as $c_m = \langle \Psi_m | \Psi_0 \rangle$, with $| \Psi_0 \rangle$ being the initial state.
\begin{figure}[!t]
    \includegraphics[width=\linewidth]{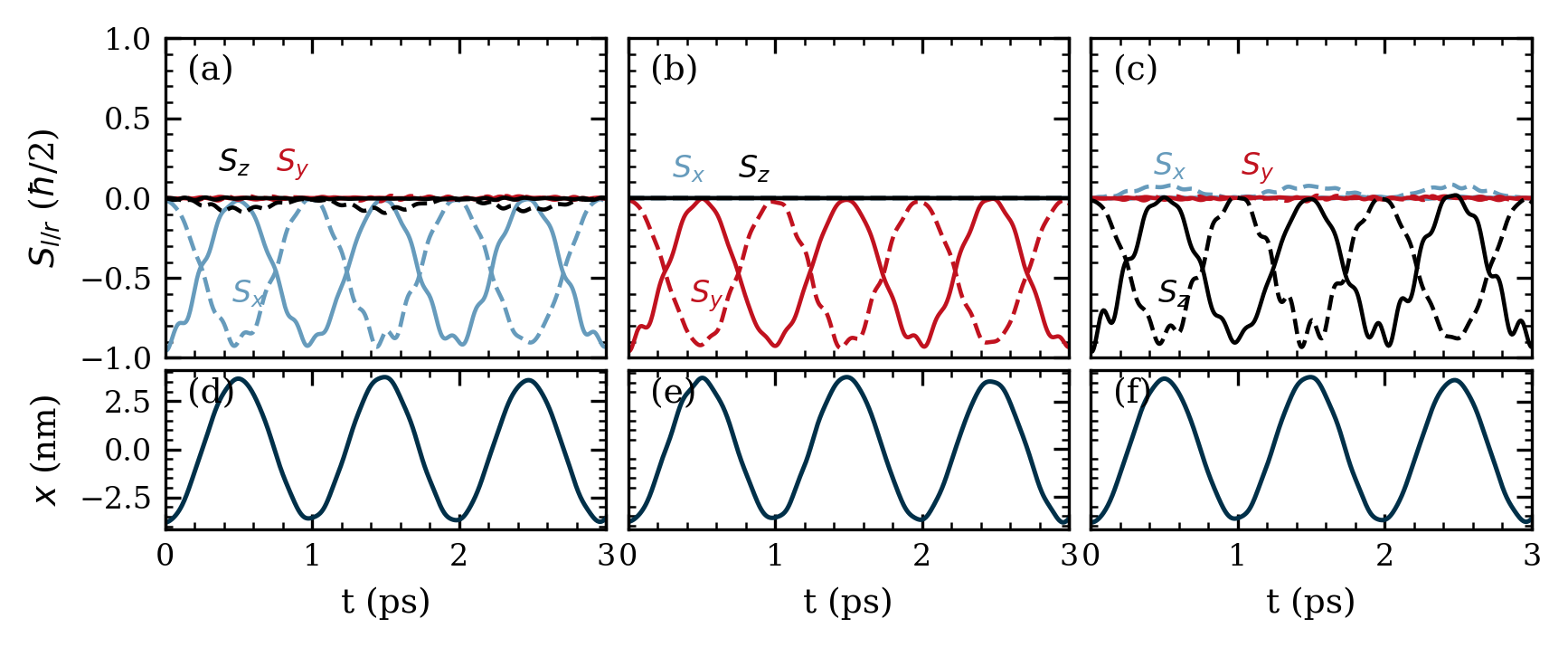}
    \caption{Time evolution of the expectation value of the spin in the left (solid lines) and right (dashed lines) dots. The $x$, $y$, and $z$ components of the spin are marked in blue, red, and black, respectively. Panels [(d)–(f)] show the average electron position $x$ as a function of time. Results for a quantum dot size of $d = 5.5~\mathrm{nm}$.}
    \label{fig:one_electron_small}
\end{figure}

\begin{figure}
    \includegraphics[width=\linewidth]{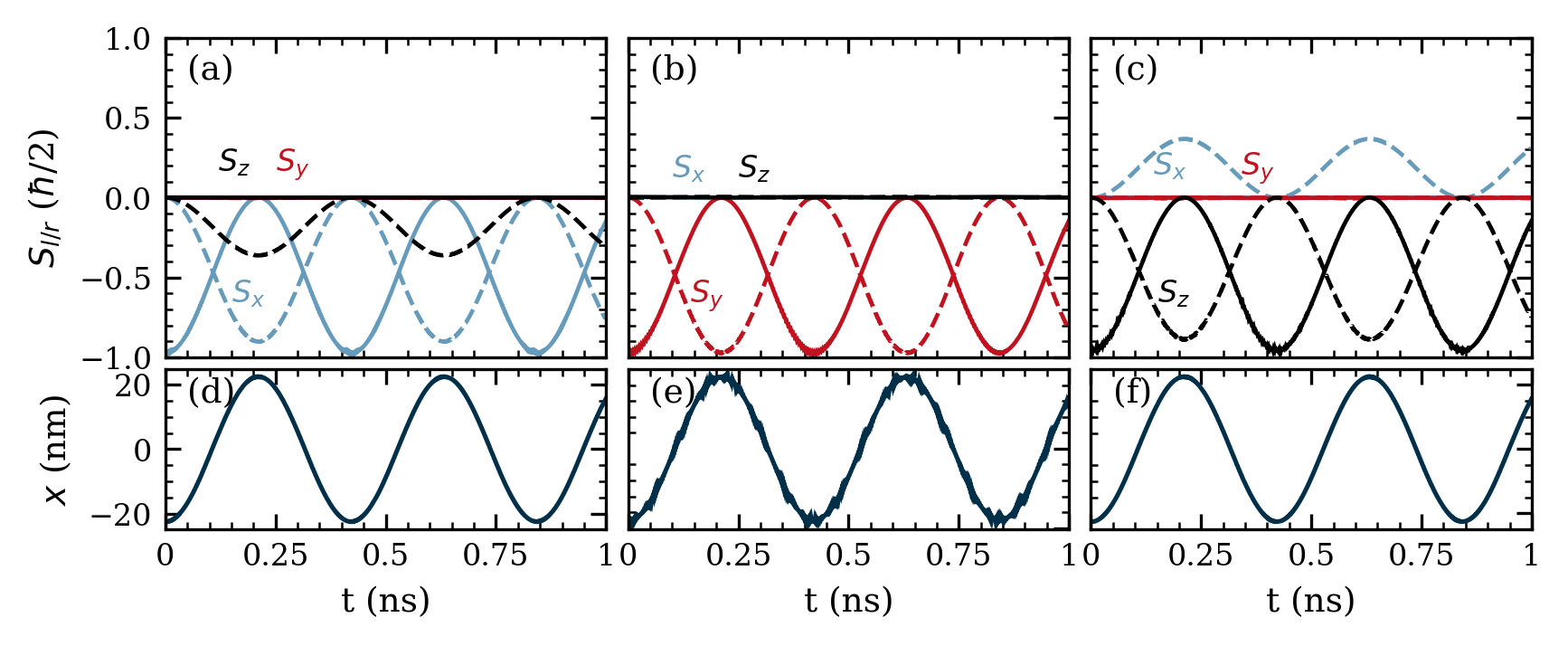}
    \caption{Same as Fig.~\ref{fig:one_electron_small}, but for a quantum dot size of $d=33$ nm.}
    \label{fig:one_electron_large}
\end{figure}

In this section, we consider a double QD with two different QD sizes, $d = 5.5~\mathrm{nm}$ and $d = 33~\mathrm{nm}$, corresponding to two distinct regimes. In the large-dot regime ($d=33$~nm), the electronic spectrum is dominated by the $d_{xy}$ states, as the level spacing induced by the confinement is significantly smaller than the energy splitting $\Delta _E$ between the $d_{xy}$ and $d_{xz/yz}$ orbitals - see Eq.~(\ref{eq:H0}). In contrast, for small dots ($d=5.5$~nm), the confinement-induced quantization becomes sufficiently strong that the $d_{xz}$ and $d_{yz}$ bands, characterized by a large effective mass along one of the direction and determined by the $t_h$ parameter [Eq.~(\ref{eq:H0})], contribute to the low-energy levels. Figure~\ref{fig:single_el_spectrum_as_size} shows the single-electron spectrum of the double QD as a function of $d$, with contributions from different orbitals indicated using an RGB color scheme. Indeed, for $d$ below $10$~nm, the lowest excited states are composed of significant contributions from the $d_{yz/xz}$ orbitals.

The time evolution of the expectation values of the spin components in the left and right quantum dots, together with the average electron position along the $x$ direction, is shown in Fig.~\ref{fig:one_electron_small} and Fig.~\ref{fig:one_electron_large} for $d=5.5~\mathrm{nm}$ and  $d=33~\mathrm{nm}$, respectively. In both cases, the electron undergoes coherent tunneling between the dots, with a period that depends on the dot size: $T \approx 1~\mathrm{ps}$ for $d=5.5~\mathrm{nm}$ and $T \approx 425~\mathrm{ps}$ for $d=33~\mathrm{nm}$. These results are consistent with the tunneling times extracted from the relation $T=\frac{2\pi\hbar}{\Delta E}$ which gives: $T=0.986~\mathrm{ps}$ for $d=5.5~\mathrm{nm}$, and $T=420~\mathrm{ps}$ for $d=33~\mathrm{nm}$

For initial spin orientations antiparallel to the $x$ and $z$ axes [Figs.~\ref{fig:one_electron_small} and \ref{fig:one_electron_large}, panels (a) and (c)], the spin acquires finite $S_z$ and $S_x$ components during the time evolution. This behavior arises from spin precession in the effective SO field ($\mathbf{B}_{SO}$) which occurs when the initial spin has a component perpendicular to $\mathbf{B}_{SO}$. For the low-energy states, predominantly determined by the $d_{xy}$ orbital, the SO coupling has a Rashba-like form \(\alpha_R [\sigma_y \sin(k_x a) - \sigma_x \sin(k_y a)]\), with $\alpha_R$ given by \(\alpha_R = \Delta_{\mathrm{SO}} \Delta_{\mathrm{RSO}} / 3\Delta _E\) (see Appendix~\ref{sec:a1} for details). In this case  an effective SO field $\mathbf{B}_{SO}$ is oriented perpendicular to the electron momentum. As the electron tunnels between the dots, the corresponding momentum along the $x$ direction induces an effective Rashba field oriented along the $y$ axis, which changes sign when the electron reverses its direction of motion.  Note that a larger double QD is characterized by a longer tunneling period. Consequently, the time over which the spin can undergo coherent precession about   $\mathbf{B}_{\mathrm{SO}}$, up to momentum reversal, is extended. This results in larger-amplitude oscillations of $S_x$ and $S_z$ compared to the small-dot case.
Consistently, no acquisition of the $S_x$ and $S_z$ components (no precession) is observed when the spin is initially aligned along the $y$ axis [Figs.~\ref{fig:one_electron_small} and \ref{fig:one_electron_large}, panels (b)], confirming that the effective SO field is predominantly oriented along this direction.

\begin{figure}
    \centering
    \includegraphics[width=\linewidth]{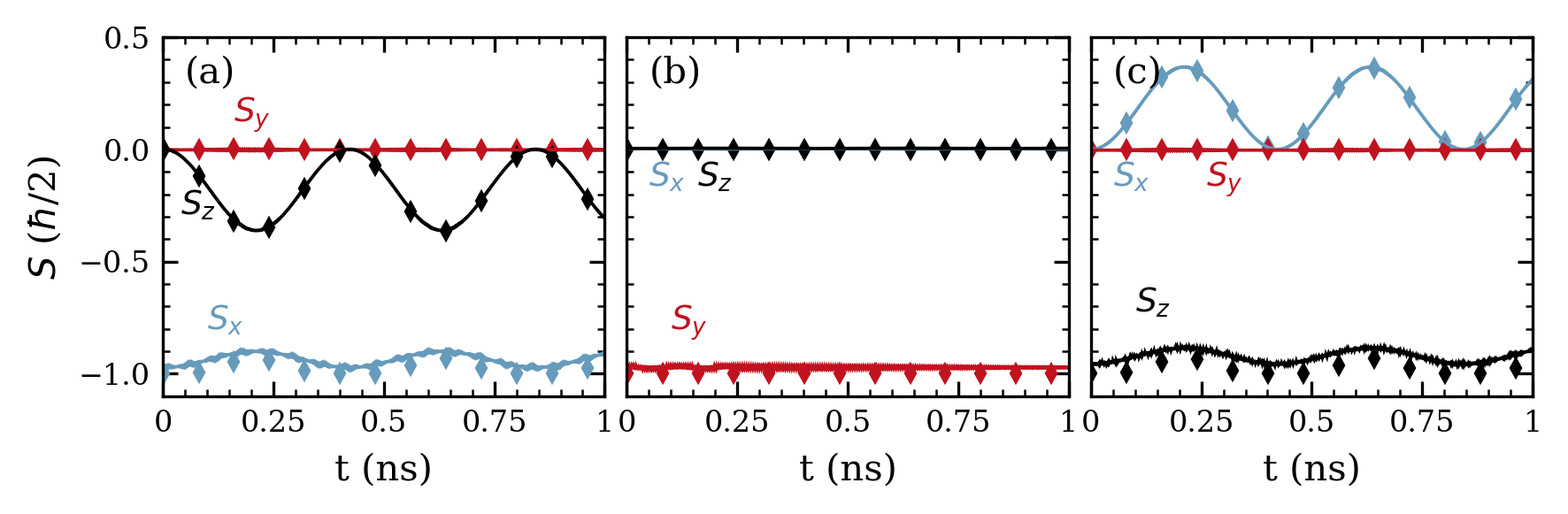}
    \caption{The expectation value of spin from the full quantum mechanical calulations presented in Fig.~\ref{fig:one_electron_large} (solid lines) compared with the solution of the Bloch equations with a Rashba-like SO coupling, Eq.~(\ref{eq:bloch}) (indicated by diamonds). Results for $d=33$~nm.}
    \label{fig:bloch_large}
\end{figure}

\begin{figure}
    \centering
    \includegraphics[width=\linewidth]{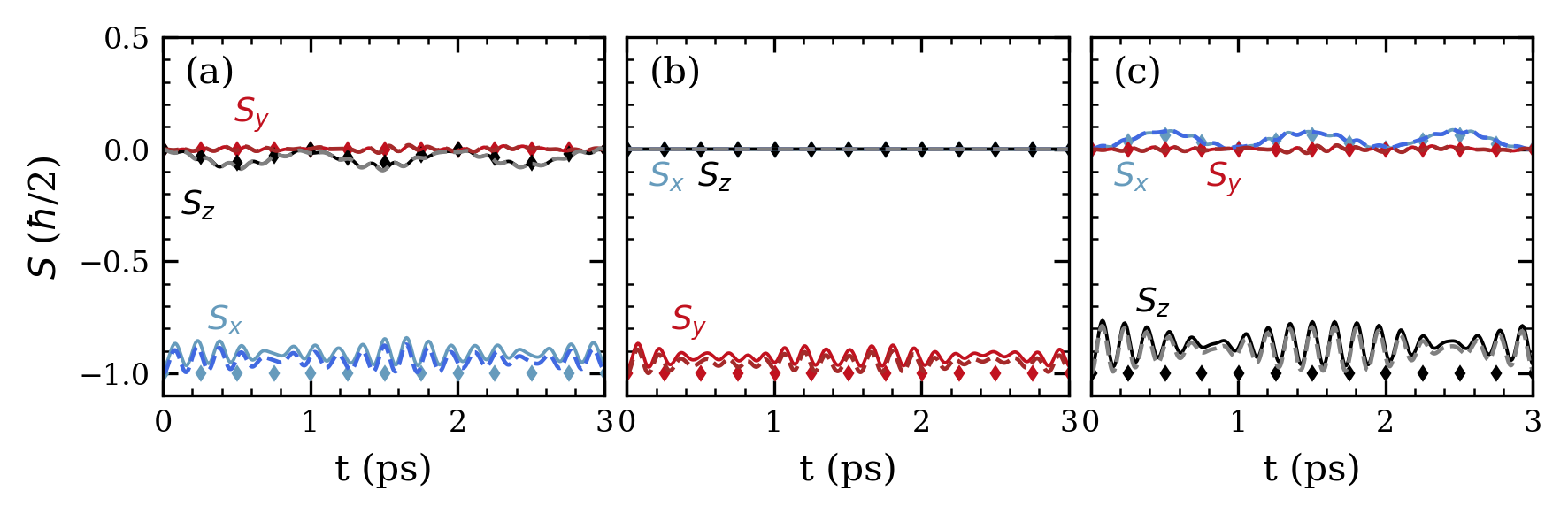}
    \caption{The expectation value of the spin from the full quantum mechanical calulations presented in Fig.~\ref{fig:one_electron_small} (solid lines) compared with the solution of the Bloch equations with a SO coupling in the Rashba form, Eq.~(\ref{eq:bloch}) (marked by diamonds) as well as the solution of the Bloch equation with the coupling between the spin and orbital angular momentum, Eq.~(\ref{eq:bloch2}) (dashed lines). Results for $d=5.5$~nm.}
    \label{fig:bloch_small}
\end{figure}

It is noteworthy that for $d = 5.5~\mathrm{nm}$ [Fig.~\ref{fig:one_electron_small}], the spin dynamics display less regular oscillatory behavior, characterized by a pronounced beating pattern. This behavior arises from the contribution of higher excited states, which for such a small QD size have $d_{xz/yz}$ character - see Fig.~\ref{fig:single_el_spectrum_as_size}. In this case, the SO coupling deviates from the simple Rashba-like form due to the presence of strong interorbital coupling, which leads to more complex spin dynamics and gives rise to the beating structure observed in Fig.~\ref{fig:one_electron_small}.\\

% OPIS BLOCHA
Our conclusions concerning the spin dynamics of an electron in a double QD can be additionally supported by a semi-classical approach based on the Bloch equation
\begin{equation}
    \frac{d \mathbf{S} }{dt} = \frac{g \mu_B}{\hbar}  \mathbf{S} \times \mathbf{B}_{SO},
 %   \label{eq:bloch}
\end{equation}
where in the first approximation,  $\mathbf{B}_{SO}$ is assumed to have a Rashba-like form and is given by
\begin{equation}
\mathbf{B}_{\mathrm{SO}} = \frac{2 \alpha_R}{g \mu_B} [k_y, -k_x, 0]^T, 
\end{equation}
where \(\alpha_R = \Delta_{\mathrm{SO}} \Delta_{\mathrm{RSO}} / 3\Delta E\).
According to the analysis presented in Appendix~\ref{sec:a1}, the assumed simplified form of the SO coupling in the LAO/STO interface is valid for the $d_{xy}$ orbitals and remains applicable in the large-dot regime. Finally, the Bloch equation takes the form
\begin{equation}
\left (
\begin{array}{c}
\frac{dS_x}{dt} \\
\frac{dS_y}{dt} \\
\frac{dS_z}{dt}
\end{array} 
\right ) = 
\frac{g \mu_B}{\hbar}
\left (
\begin{array}{c}
- k_x S_z  \\
- k_y S_z  \\
\ k_x S_x + k_y S_y 
\end{array} 
\right ),
\label{eq:bloch}
\end{equation}
where, in the calculations, the wave vector is treated in the semiclassical sense as $k_{x(y)} = m V_{x(y)} / \hbar$, where $V_{x(y)} = \frac{d\langle x \rangle }{dt} \left ( \frac{d \langle y \rangle}{dt}\right )$ and $m = \frac{ \hbar }{ 2 a^2 t_l }=0.28$ is the effective mass corresponding to the $d_{xy}$ orbital.
 
Figure~\ref{fig:bloch_large} presents a comparison of the averaged spin components from the full quantum mechanical calculations (solid lines) with those obtained from the Bloch equation (marked by diamonds), for $d=33$~nm. As the Bloch approach does not allow for resolving the spin in separate QDs, the figure shows the spin evaluated over the entire system. We can see that the amplitude and period of the spin-component oscillations in time are well reproduced by the Bloch equation with the Rashba-like SO field for the considered large quantum dots ($d=33$~nm). Here, the slight difference in the amplitudes arises from the initial conditions. The semiclassical approach assumes a separation of spin and positional degrees of freedom, such that the spinor remains a pure state on the Bloch sphere with $\langle S \rangle=1$. In contrast, in the full quantum calculation, the expectation value of the spin for the chosen initial condition is not exactly 1 due to the presence of SO coupling.

Note that, in the considered large-dot regime, the spin dynamics can also be accurately described in terms of the Bloch equations. For this purpose, let us assume that the electron spin is initially oriented antiparallel to the $x$ or $z$ direction [Fig.~\ref{fig:bloch_large} (a,c)],  perpendicular to the effective $\mathbf{B}_{\mathrm{SO}}$ field. From Eq.~(\ref{eq:bloch}), a nonzero $S_x$ ($S_z$) combined with
$k_x \neq 0$ leads to a change in the $S_z$ ($S_x$) component, which acquires the opposite sign for the two initial spin polarizations, as presented in Fig.~\ref{fig:bloch_large}(a, c). In time, the nonzero $S_z$ ($S_x$) generates the acquisition of the $S_x$ ($S_z$) component through the first and third equations in Eq.~(\ref{eq:bloch}), leading to oscillations of both components in phase and antiphase for the initial spin polarization along the $x$ and $z$ axes, respectively; see Fig.~\ref{fig:bloch_large}(a,c). On the other hand, when the initial spin is polarized along the $\mathbf{B}_{\mathrm{SO}}$ field, the $y$ axis, no spin precession occurs [Fig.~\ref{fig:bloch_small} (b)], since in Eq.~(\ref{eq:bloch}) the left-hand side is nearly zero due to $\langle k_y \rangle(t)=0$, induced by the strong confinement in the $y$ direction.

The agreement between the quantum-mechanical calculations and the Bloch equation with Rashba-like SO coupling breaks down in the small-dot regime, as presented in Fig.~\ref{fig:bloch_small}. In this case, a beating pattern for all initial spin polarizations is observed, associated with the participation of the $d_{yz}$ and $d_{xz}$ orbitals. In this regime, Eq.~(\ref{eq:bloch}) predicts no spin precession, as the electron tunneling between the dots is so fast that the average value of $k_x$ approaches zero, corresponding to a highly nonadiabatic regime.
In this case, the observed spin dynamics originates from interorbital coupling governed by both $H_{\mathrm{SO}}$ and $H_{\mathrm{RSO}}$, reflecting the coupling between the spin and the orbital angular momentum. In this case, the Bloch equation describing the spin dynamics takes the form
\begin{equation}
\frac{d \mathbf{S}}{dt} = \frac{2 \Delta_{\mathrm{SO}}}{3\hbar} \langle \mathbf{S} \times \mathbf{L} \rangle,
\label{eq:bloch2}
\end{equation}
and depends on the $\mathbf{S} \times \mathbf{L}$ coupling averaged over quantum state.

In Fig.~\ref{fig:bloch_small}, the dashed lines depict the results obtained from calculations based on Eq.~(\ref{eq:bloch2}). This equation accurately reproduces the results of the full quantum-mechanical calculations, including the characteristic beating pattern of $S_z$.

\subsection{SWAP operation}
\label{sec:swap}
Now, we consider two electrons confined in a double QD based on the LAO/STO 2DEG and analyze the SWAP operation. For this purpose, as an initial condition we assume that the electrons are spatially localized in two distinct quantum dots (left and right) and their spin are prepared in an antiparallel configuration. The initial two-electron state is taken as the antisymmetrized product
$\Psi_{0}(\mathbf{r}_1,\mathbf{r}_1) = \frac{1}{\sqrt{2}} \left( \Psi_l(\mathbf{r}_1) \Psi_r(\mathbf{r}_2) - \Psi_l(\mathbf{r}_2) \Psi_r(\mathbf{r}_1) \right)$,
where the single-electron states, localized in the left (right) dot $\Psi_l$ ($\Psi_r$), are constructed as linear combinations of bonding and antibonding eigenstates of the single-electron Hamiltonian (\ref{eq:Hamiltonian_real_space}). Note that $\Psi_l$ and $\Psi_r$ are evaluated under the assumption of a weak external magnetic field of magnitude $10^{-5}\,\mathrm{T}$ to polarize the electron spins along a chosen axis.

The initial state is projected onto the eigenstates of the two-electron Hamiltonian [Eq.~(\ref{eq:two_electrons})], determined by the CI method,  to obtain the coefficients $c_m = \langle \Psi_m \mid \Psi_0 \rangle$, which are then used to determine the time evolution according to Eq.~(\ref{eq:evolution}). Consequently, the spin dynamics depend on the contributions of the two-electron eigenstates to the initial wave function, which may have different orbital characters and can change when the size of QD is varied.
\begin{figure}
    \centering
    \includegraphics[width=0.8\linewidth]{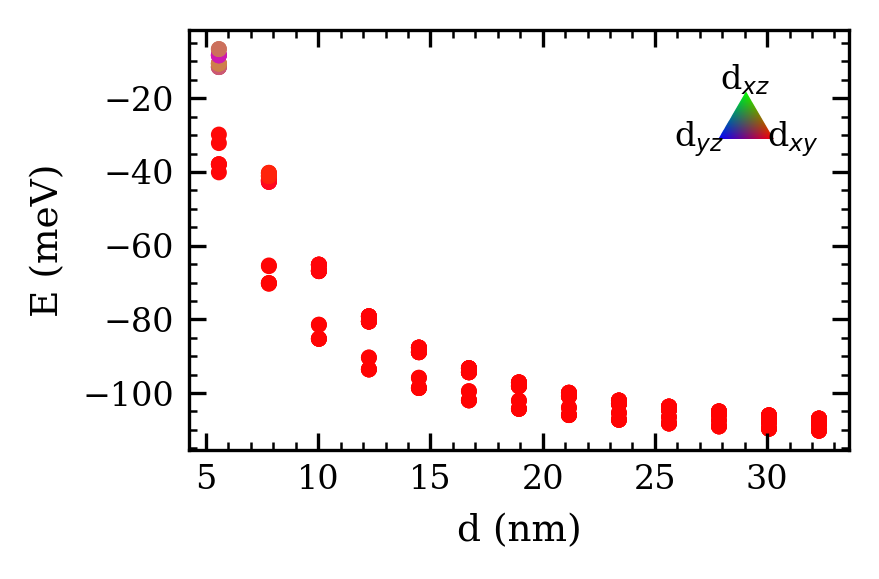}
    \caption{Two-electron energy spectrum of a double LAO/STO QD as a function of the size of the single QD. Contributions from different orbitals $d_{xy}$, $d_{xz}$ and $d_{yz}$ are indicated using an RGB color scheme. Results for the potential parameters $V_b =70$ meV and $V_0=10$ meV.}
    \label{fig:e_in_size}
\end{figure}
\begin{figure*}
    \centering
    \includegraphics[width=0.8\linewidth]{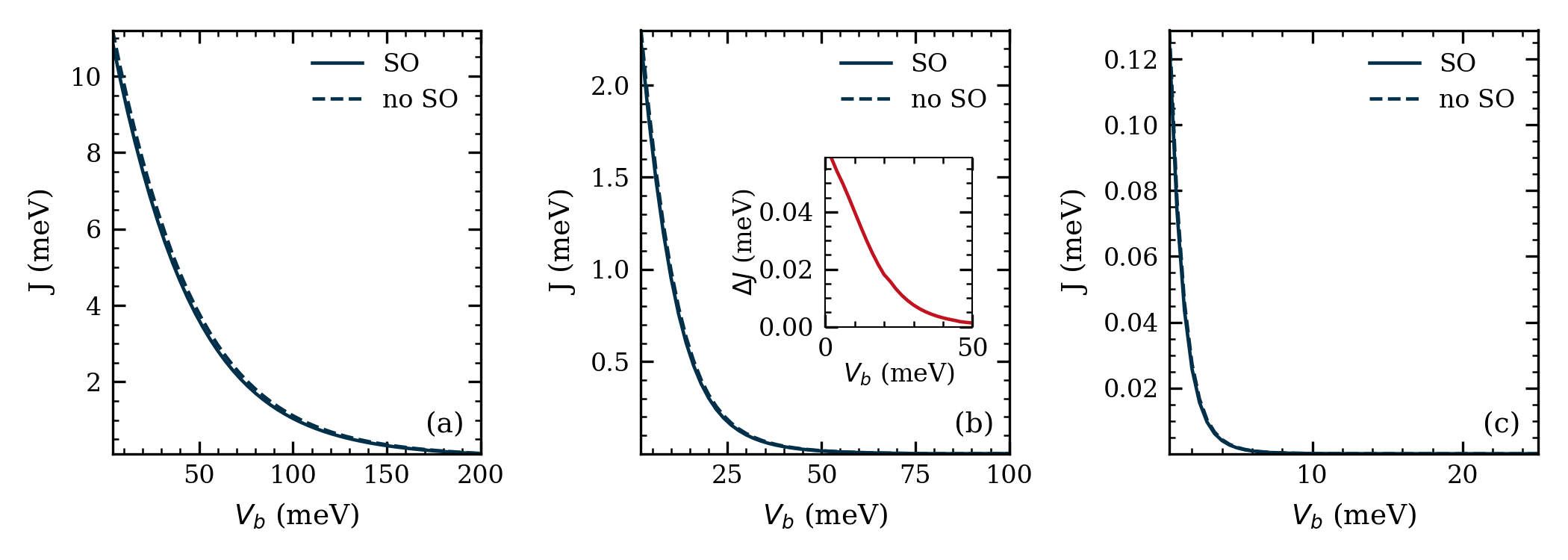}
    \caption{Exchange energy $J$ as a function of the barrier height $V_b$ separating quantum dots with diameters (a) $d = 5.50$ nm, (b) $d = 11$ nm, and (c) $d = 33$ nm. Results for a confinement potential characterized by $V_b = 70$ meV and $V_0 = 10$ meV. The inset in panel (b) presents the difference between the exchange energy calculated with in the full model, including SO coupling, and that without SO interaction. The inset in panel (c) shows the crossing of the lowest-energy singlet (in black) and triplet state (in blue), in the range of $V_b$ corresponding to the dip in $J$ at $V_b=5$~meV.}
    \label{fig:exchange_energy}
\end{figure*}

In Fig.~\ref{fig:e_in_size}, the two-electron energy spectrum of a double QD is presented as a function of the QD size, $d$.  As \(d\) increases, the orbital character of the higher-lying excited states is modified. For small QDs ($d = 5$ nm), the lower excited states exhibit an admixture of the $d_{xz}$ and $d_{yz}$ orbitals, which results from their effective-mass anisotropy, characterized by a relatively large value along one direction - see Eq.~(\ref{eq:H0}).
To study the QD size dependence of the SWAP operation, in the further part of the paper, we analyze three representative QD diameters, $d = 5.5$ nm, $d = 11$ nm, and $d = 33$ nm, for which the contribution of the $d_{xz}$/$d_{yz}$ orbitals to the lower-energy electronic states gradually diminishes.

For the chosen $d$, the exchange energy $J$, defined as the difference between the energy of the spin-singlet and spin-triplet states, is presented in Fig.~\ref{fig:exchange_energy} as a function of the barrier potential separating the quantum dots. 
The magnitude of $J$ determines the switching time of the SWAP operation according to the standard relation $t_{\mathrm{SWAP}} = \frac{\pi \hbar}{J}$. Since the barrier height, \(V_b\), determines the inter-dot coupling, the exchange energy decreases monotonically with increasing \(V_b\). Moreover, $J$ becomes larger for smaller quantum dots, where electron–electron interactions are enhanced as a consequence of the strong confinement. For comparison, in Fig.~\ref{fig:exchange_energy} we also present the exchange energy determined in the absence of SO coupling (dashed line). For $d=11$~nm [Fig.~\ref{fig:exchange_energy}(b)], the difference in $J$ determined based on the full model, including SO coupling, and that without SO interaction is shown in the inset.

\begin{figure*}
    \centering
    \includegraphics[width=0.8\linewidth]{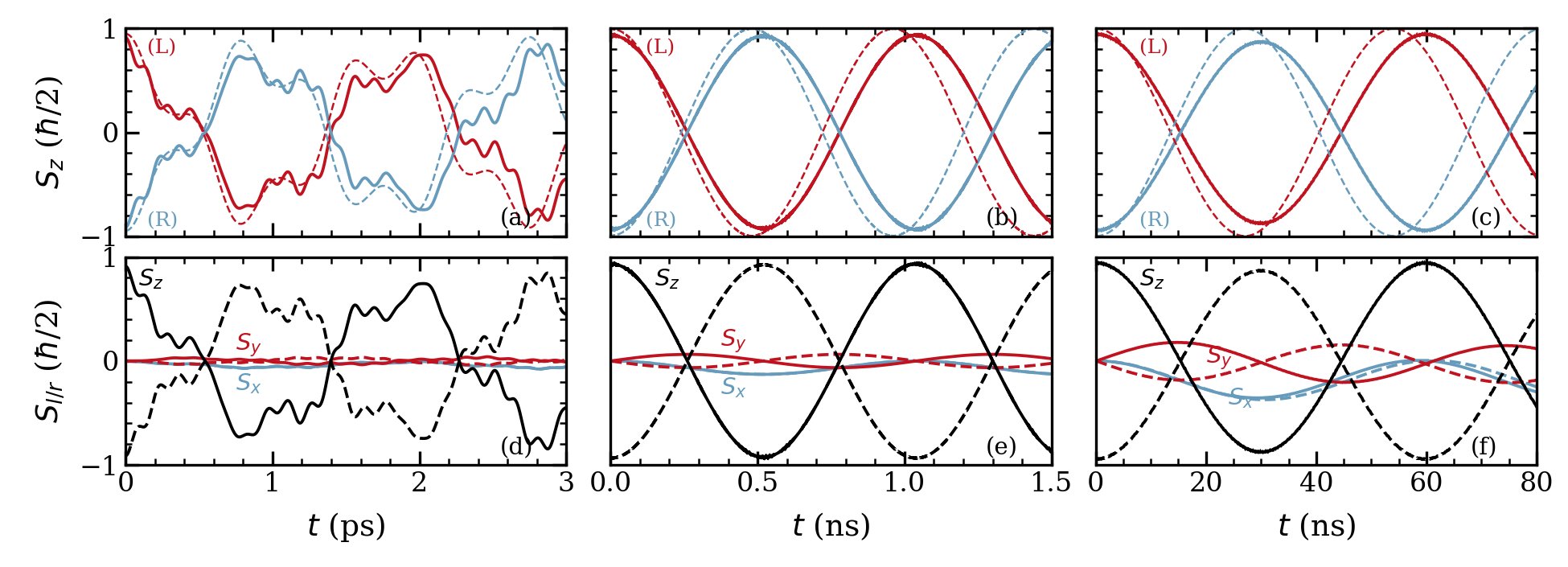}
    \caption{Time evolution of the expectation value of the spin component $S_z$ for double QD of different diameters: $d = 5.5$ nm (a, d), $d = 11$ nm (b,e), and $d = 33$ nm (c,f). Results for diameters $d = 5.50$~nm and $d = 11$~nm are obtained for the potential parameters $V_b = 70$~meV and $V_0 = 10$~meV, while for $d = 33$~nm the parameters are $V_b = 10$~meV and $V_0 = 10$~meV. The upper panels (a-c) compare the  $S_z$ evolution obtained within the model with SO (solid lines) and without SO coupling (dashed lines). The lower panels (d-f) show the time evolution of the $S_x$, $S_y$, and $S_z$ spin components in the left (solid lines) and right (dashed lines) quantum dot.}
    \label{fig:spin_evolution}
\end{figure*}
\begin{figure}
    \centering
    \includegraphics[width=\linewidth]{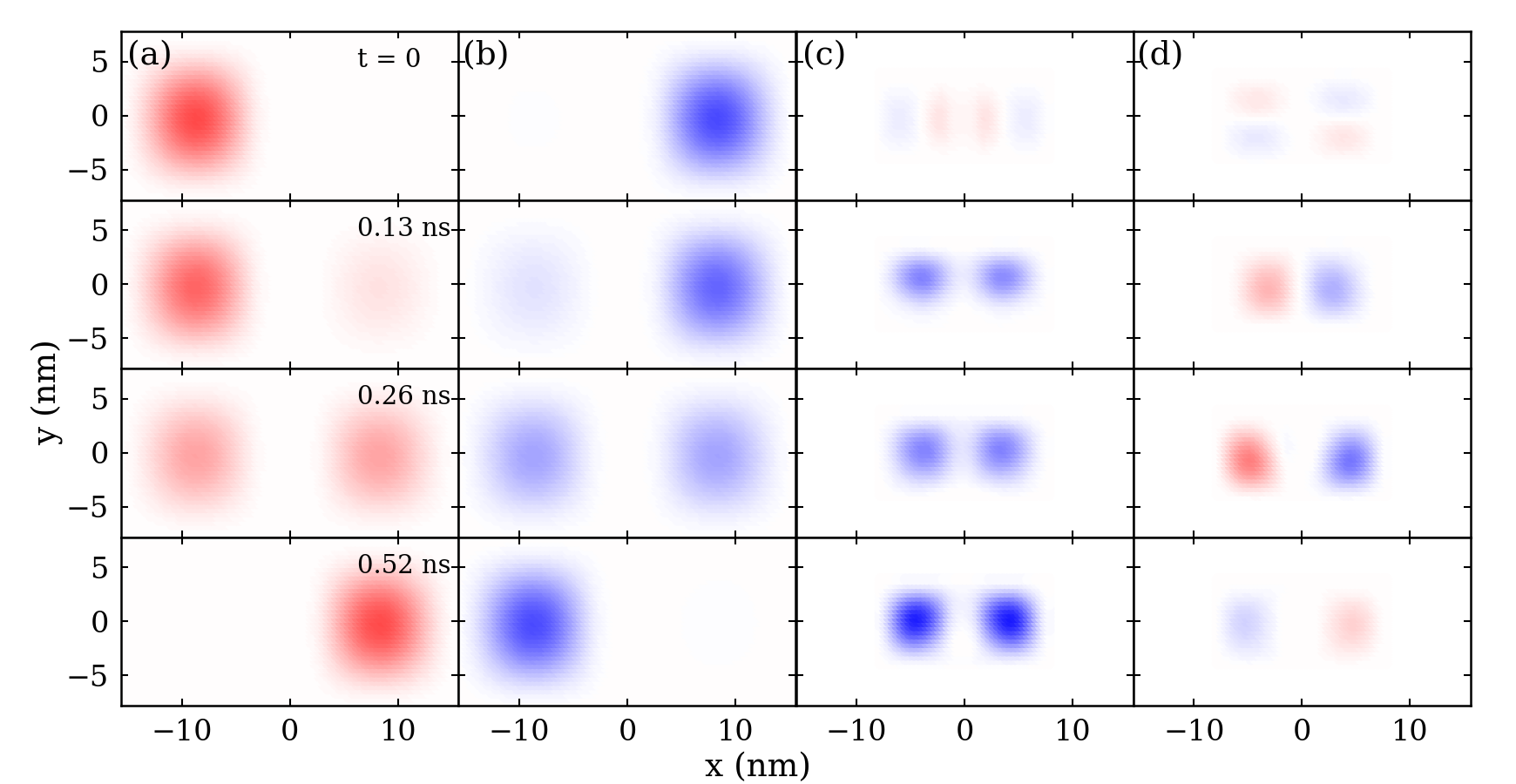}
    \caption{Spin density maps at selected time instances for a double QD with diameter $d = 11$ nm. Panel (a) and (b) correspond to $s_z$; separately, for spin up (a) and down (b) components, while (c) and (d) show the $s_x$ and $s_y$ spin density, respectively.}
    \label{fig:spinmap-medium}
\end{figure}
In Fig.~\ref{fig:spin_evolution}(a–c) we present the time evolution of the spin expectation values $S_z$ in the left (red) and right (blue) QD for two electrons confined in separate QDs, initially prepared with opposite spin oriented along the $z$ axis. The corresponding results obtained within a model without SO coupling are shown by dashed lines. Note that at $t=0$ the spins in the left and right dots are not exactly equal to $\pm \hbar/2$ due to the SO coupling and leakage of $\Psi_l$ and $\Psi_r$ functions to the right and left dot.  For the QD diameters $d$ under consideration, we observe an exchange of the electron spin between the dots with a period $t_{\mathrm{SWAP}} \approx 1$~ps for $d = 5.5$~nm, $t_{\mathrm{SWAP}} \approx 0.5$~ns for $d = 11$~nm, and $t_{\mathrm{SWAP}} \approx 29$~ns for $d = 33$~nm. The corresponding switching times, determined from the relation $t_{\mathrm{SWAP}} = \frac{\pi \hbar}{J}$, with $J$ extracted from the two-electron energy spectrum (Fig.~\ref{fig:e_in_size}), are equal to $0.94$~ps, $0.51$~ns, and $28.94$~ns, respectively. 

Note that only for the moderate-QD size, $d=11$~nm [Fig.~\ref{fig:spin_evolution}(b)], the dynamics underlying the SWAP operation exhibit coherent oscillations that enable an almost complete transfer of spin polarization between the dots. The time evolution of all spin components ($S_x$, $S_y$, and $S_z$) is displayed in Fig.~\ref{fig:spin_evolution}(d–f), where the spin dynamics in the left and right QDs are distinguished by solid and dashed lines, respectively. Note that for $d=11$~nm, the impact of the SO interaction, detrimental to the spin exchange, is strongly suppressed by the confinement in the $x$ direction. In Fig.~\ref{fig:spin_evolution}(e), only minor oscillations of the transverse spin components $S_x$ and $S_y$ are observed. For this specific value of $d$ in Fig.~\ref{fig:spinmap-medium}, we present the spin density at selected time instants during the SWAP operation. The exchange of the electronic spin component $S_z$ is clearly visible in Fig.~\ref{fig:spinmap-medium}(a,b), where the spin density $s_z$, determined separately for the spin-up and spin-down wave function components, is shown.
\begin{figure}
    \centering
    \includegraphics[width=\linewidth]{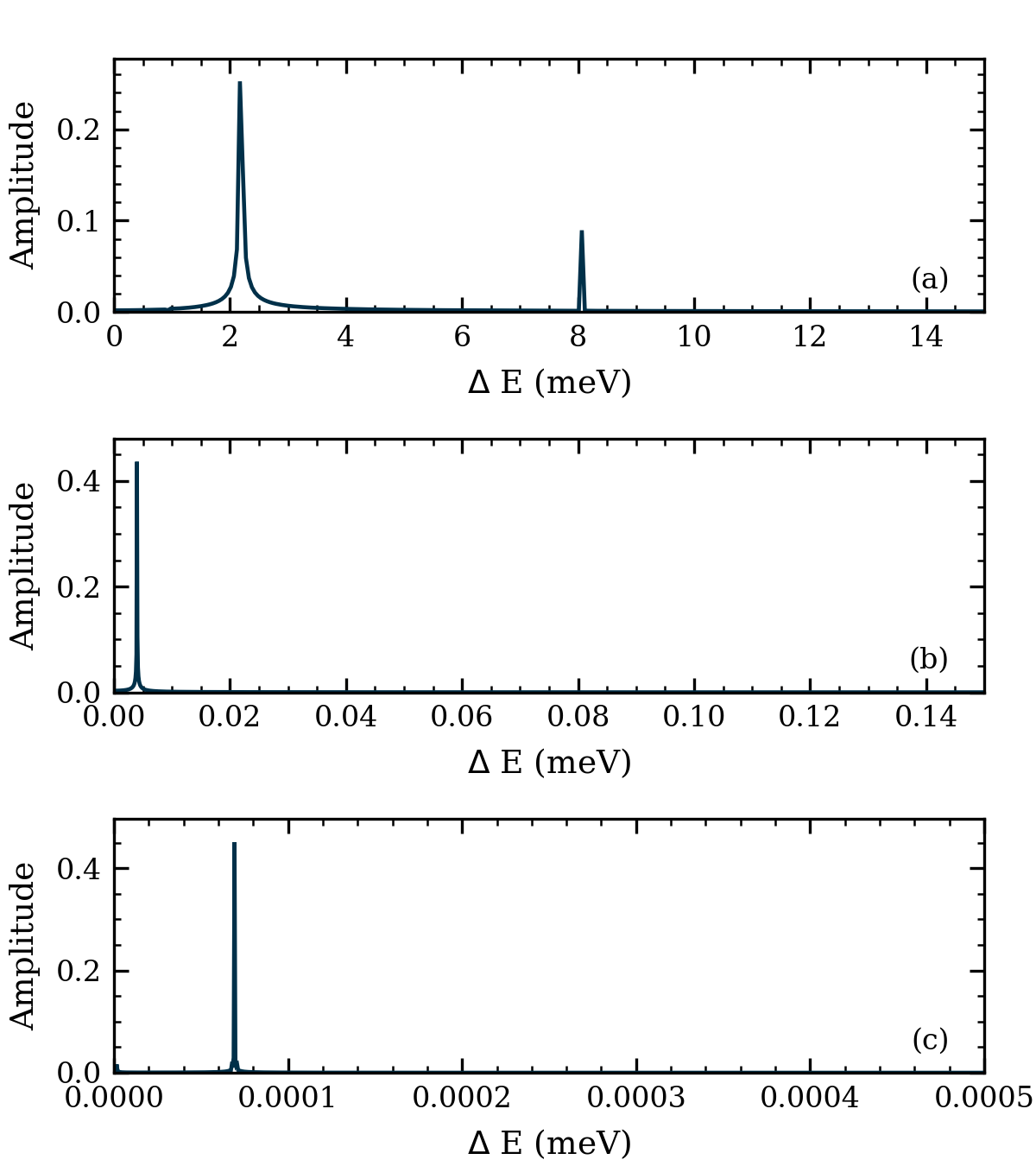}
    \caption{Fourier spectra of the time evolution of expectation value of  $S_z$ component presented in Fig.~\ref{fig:spin_evolution}. Results for QD with diameter (a) $d=5.5$ nm, (b) $d=11$ nm, (c) $d=33$ nm.}
    \label{fig:fourier}
\end{figure}
\begin{table}
\centering
\caption{The coefficients $C_{nm} = c_n^* c_m$, the energy differences $\Delta E_{nm}$, and the matrix elements $\langle n | S_z | m \rangle$ are determined for pairs of two-electron eigenstates of double quantum dot systems with different diameters.}
\vspace{0.2cm}
\begin{ruledtabular}
\renewcommand{\arraystretch}{1.15}
\begin{tabular}{c c c l}
$d$ [nm] & $(n,m)$ & $C_{nm}$ & $\Delta E_{nm}$ [meV] \\
\hline
 & $(1,3)$ & 0.3979 & 2.190 \\
5.50 & $(3,6)$ & 0.2084 & 8.065 \\
 & $(1,6)$ & 0.1846 & 10.256 \\
\hline
11 & $(1,3)$ & 0.4809 & 0.0039 \\
\hline
33 & $(1,3)$ & 0.4851 & $6.92\times10^{-5}$ \\
\end{tabular}
\end{ruledtabular}
\label{table}
\end{table}

When the quantum dot size is increased to $d = 33~\mathrm{nm}$ [Fig.~\ref{fig:spin_evolution}(c,f)], the SO energy becomes larger relative to the confinement/kinetic energy, disturbing the SWAP operation. In a semiclassical picture, the SO coupling leads to an additional spin precession about the effective SO field. 
%Note that for large QD sizes, the electronic spectrum is fully determined by the $d_{xy}$ orbital [Fig.~\ref{fig:e_in_size}] with the SO coupling of Rashba type. 
Figure~\ref{fig:spin_evolution}(f) shows that the amplitude of the initial spin state with $S_z \approx \pm \hbar/2$ is reduced during the time evolution (i.e., during the subsequent SWAP operation). This process is accompanied by a significant growth of the transverse spin components $S_x$ and $S_y$, whose amplitudes increase in time as a result of spin precession around the effective SO field. Note in Fig.~\ref{fig:spin_evolution}(c) that the leakage of $S_z$ is absent in the model without SO interaction (dashed lines). In this case, the components \(S_x\) and \(S_y\) remain zero throughout the spin evolution (not shown here), and the switching time is reduced due to the sightly larger exchange energy as presented in Fig.~\ref{fig:exchange_energy}.

For QDs with a small diameter of $d = 5.5$~nm [Fig.~\ref{fig:spin_evolution}(a,d)], the exchange spin dynamics are strongly perturbed and exhibit an irregular shape. This irregular evolution is only partially attributable to the SO interaction, as indicated by the dashed lines in Fig.~\ref{fig:spin_evolution}(a), showing results without SO coupling. The dominant mechanism in this case arises from electronic transitions involving higher-lying excited states with a substantial contribution from the \(d_{xz}\) and \(d_{yz}\) orbitals.

To distinguish between the pair of electronic states that participate in the SWAP operation, we first perform a Fourier transform on the time evolution $S_z(t)$, shown in Fig.~\ref{fig:fourier}. The expectation value of the spin is determined by 
\begin{equation}
S_i=\langle \Psi(t) | \hat{S}_i | \Psi(t) \rangle 
= \sum_{n,m} c_m c_n^* \, e^{\frac{i}{\hbar}(E_n - E_m)t} \, \langle \Psi_n | \hat{S}_i | \Psi_m \rangle,
\label{eq:spin}
\end{equation}
where $i=x,y,z$, and it depends on the contributions of the two-electron eigenstates $n$ and $m$ to the initial state through the factor $C_{nm} = c_n^* c_m$. The expectation value $\langle \Psi_n | \hat{S}_i | \Psi_m \rangle$ oscillates with the frequency defined by the energy difference $ \Delta E_{nm} = E_n - E_m$. For three selected quantum dot QD sizes, in Table~\ref{table} we present the coefficients $C_{nm}$ in descending order, together with the corresponding energy differences $\Delta E_{nm}$. Note that the Hamiltonian in Eq.~(\ref{eq:Hamiltonian_k_space}) commutes with a generalized diagonal parity operator
\begin{equation}
\hat{\Pi} = \mathrm{diag}[P, -P, -P, P, -P, P],    
\label{eq:parity}
\end{equation}
where \(P\) is the scalar parity operator defined as $P \psi(\mathbf{r}) = \psi(-\mathbf{r})$.
As a result, each component of the eigenfunctions has a definite scalar parity, either even or odd, and the matrix elements of $\hat{S}_i$ integrated over the left or right QD are nonzero only when calculated between states with opposite parity. Moreover, it is worth noting that in the presence of SO coupling, the spin is not a good quantum number, and the spin-singlet $S$ and spin-triplet states \(T_0\) and \(T_{\pm}\) are no longer purely defined. Nevertheless, for weak to moderate SO coupling strength, as in the LAO/STO 2DEG, these states have nearly-defined spin and are commonly referred to as singlet and triplet.

The Fourier analysis presented in Fig.~\ref{fig:fourier}(a) demonstrates that the spectrum of a small QD exhibits two peaks at energies corresponding to energy differences between the appropriate eigenstates. As can be seen in Table~\ref{table}, for $d = 5.5$~nm, the SWAP dynamics are governed mainly by  the transitions (1,3), corresponding to the singlet (state 1) and the triplet $T_0$ (state 3), which have opposite parity. This transition yields low-frequency oscillations in the SWAP operation, appearing as a main peak in Fig.~\ref{fig:fourier}(a). The same type of transition between the singlet and unpolarized triplet states of opposite parity, (3,6), produces a high-frequency component visible as a second peak in the Fourier spectrum. Remarkably, the higher-lying state 6 exhibits a substantial admixture of the $d_{xz/yz}$ orbitals, which disturbs the SWAP operation. Note that although the factor $C_{1,6}$, related to the transition between the ground and first excited singlet states, is sizable (Table~\ref{table}), it does not contribute to $S_z$, as both states have the same parity.

For intermediate and large dot sizes $d = 11$~nm and $d = 33$~nm, only a single prominent peak associated with the transition (1,3) is observed, consistent with the regular oscillations in Fig.~\ref{fig:spin_evolution}(e)(f). It should be noted, however, that in these cases, characterized by an increased energy of SO coupling relative to the kinetic component, the transitions generating the SWAP dynamics involve states with undefined spin character due to the presence of SO interaction. These states are characterized by non-zero spin components $S_x$ and $S_y$, whose magnitudes increase with stronger SO coupling ($d=33$~nm). As a consequence, oscillations in the $S_x$ and $S_y$ components emerge, which reduce the total fidelity of the SWAP operation, particularly for large QDs where the role of SO coupling becomes dominant.

\subsection{Anisotropy of SWAP}
\label{sec:swap_aniso}
\begin{figure}
    \centering
    \includegraphics[width=1.0\linewidth]{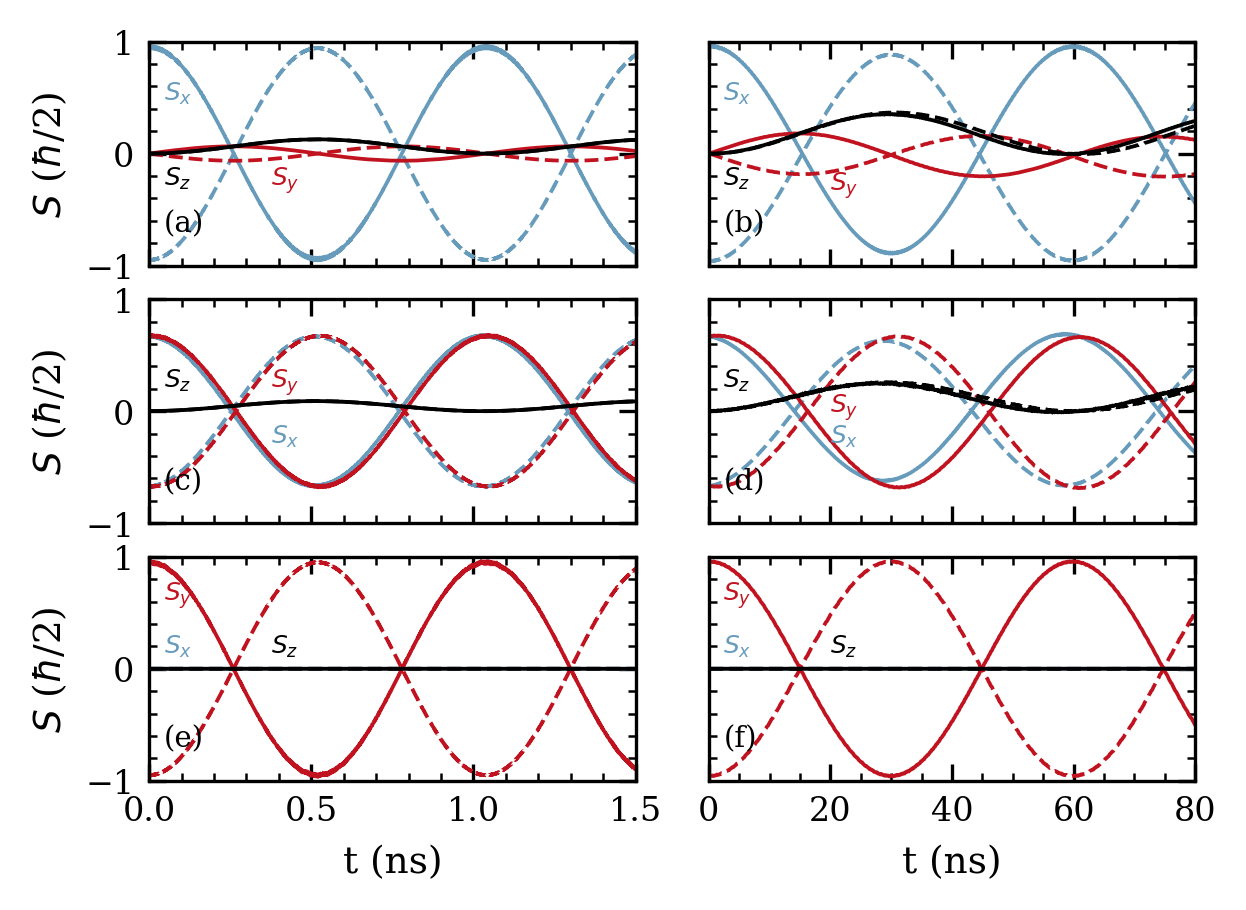}
    \caption{Time evolution of the expectation values of the spin components \(S_x\), \(S_y\), and \(S_z\) in the left (solid) and right (dashed) QD for initial spin polarization of electrons along (a,b) the \(x\)-axis, (c,d) the \(y=x\) axis, and (e,f) the \(y\)-axis. The left and right columns correspond to QDs with diameters (a,c,e) \(d = 11\)~nm and (b,d,f) \(d = 33\)~nm, respectively. To polarize the inital states, we use a low magnetic field of the order of \(10^{-5}\)~T. }
    \label{fig:aniso_evolution}
\end{figure}
We now analyze the anisotropy of the SWAP operation induced by the SO interaction. For this purpose, we consider an initial spin state lying in-plane and rotated by an angle $\theta$, where $\theta=0^\circ$ corresponds to the spin polarization along the $x$-axis ($S_x$) and $\theta=90^\circ$ is related to polarization along the $y$-axis ($S_y$).

In Fig.~\ref{fig:aniso_evolution}, we show the time evolution of the spin expectation values $S_x,S_y$ and $S_z$ in the left and right QD for three initial spin polarizations: along $x$, along the diagonal $y=x$, and along $y$. We consider QDs with $d=11$~nm and $d=33$~nm, for which the exchange of $S_z$ with high fidelity has been predicted in the previous section.
\begin{figure}
    \centering
    \includegraphics[width=0.8\linewidth]{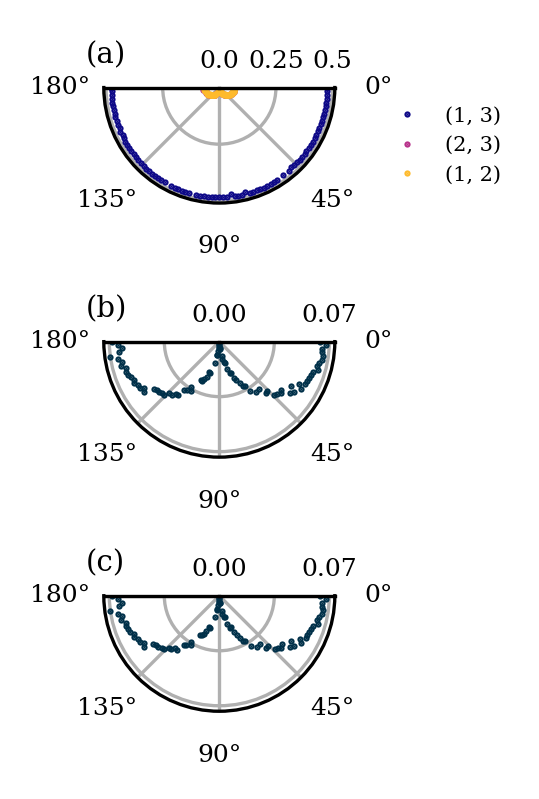}
    \caption{(a) Coefficients $C_{nm}$ as a function of the initial spin polarization angle $\theta$. Zoomed view of the anisotropy of $C_{2,3}$ (b) and $C_{1,2}$ (c). The results for a QD with diameter $d=33$~nm.}
    \label{fig:cm_large}
\end{figure}

For the first two initial spin polarizations [Fig.~\ref{fig:aniso_evolution}(a--d)], the SWAP operation is slightly perturbed by residual spin-component oscillations arising from SO coupling. In a semiclassical picture, this corresponds to spin precession about the effective SO field. For the considered dot sizes, where the electronic spectrum is predominantly determined by the $d_{xy}$ orbital, the SO field has a Rashba-like character and is directed along the $y$-axis. As a result, an initial spin polarization with an $x$ component leads to oscillations in $S_y$ and $S_z$.

Importantly, when the initial spin is aligned with the effective SO field ($y$-axis), no noticeable precession of the $S_x$ and $S_z$ spin components is observed, resulting in a nearly perfect SWAP of the electron spin [Fig.~\ref{fig:aniso_evolution}(e,f)], which is needed to design efficiently working quantum gates. The preferred direction of the initial polarization can also be understood based on Eq.~(\ref{eq:spin}) and the corresponding analysis of the transitions between eigenstates, with strengths characterized by the coefficients $C_{nm}$ illustrated in Fig.~\ref{fig:cm_large}(a--c) as a function of the polarization angle, for $d=33$~nm. As we see, the main contribution to the SWAP operation again arises from the transition $(1,3)$ between the singlet and triplet $T_0$ states of opposite parity, which is required to ensure nonzero matrix elements of the $\hat{S}_i$ in both dots. Note that the main pair $(1,3)$ is accompanied by additional transitions to the polarized triplet state $T_-$, i.e., $(2,3)$ and $(1,2)$, exhibiting strong anisotropy with respect to the initial spin polarization. Importantly, for the initial spin aligned along the $y$-axis, for which the SWAP operation is nearly perfect, the coefficients $C_{2,3}$ and $C_{1,2}$ vanish leading to high fidelity of SWAP.

\subsection{Scaled tight-binding model}
\label{sec:swap_scale}

\begin{figure}
    \centering
    \includegraphics[width=1.0\linewidth]{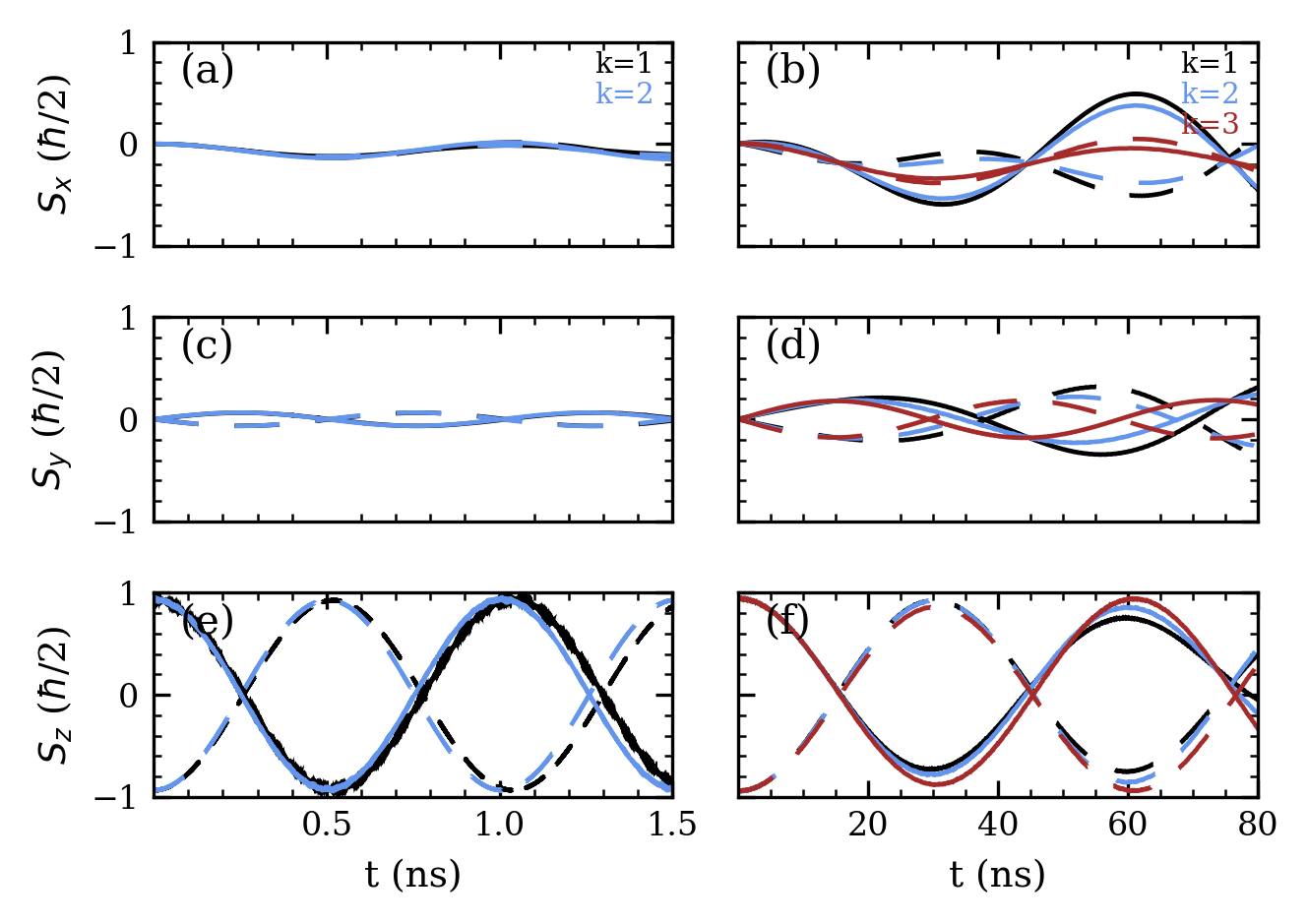}
    \caption{Comparison of the original (black lines) and the scaled (blue and red lines) model in time evolution of the (a,b) $S_x$ (c,d) $S_y$ (e,f) $S_z$ component of the spin for the dot with the diameter $d=11$~nm (left column) $d=33$~nm (right column). Solid (dashed) lines mark the spin expectation value in left (right) dot.}
    \label{fig:scaled}
\end{figure}

Since simulations of realistically sized LAO/STO-based quantum devices remain challenging due to the tight-binding Hamiltonian being defined on a square lattice with a small grid spacing $a = 0.395$~nm, in the last section we analyze the possibility of employing a scaled Hamiltonian for LAO/STO to model quantum gates, using the SWAP operation as an example. For this purpose, we adopt the model introduced in our previous paper~\cite{wojcik2025scaled}, in which a scaled tight-binding Hamiltonian was derived on a lattice with an effective spacing $dx = k a$, where $k$ is the scaling factor. In this approach, the model is mapped onto the original one by rescaling the parameters as $t'_l = t_l/k^2$, $t'_h = t_h/k^2$, $t'_d = t_d/k^2$, and $\Delta'_{\mathrm{RSO}} = \Delta_{\mathrm{RSO}}/k$, while $\hat{H}_{\mathrm{SO}}$ and $\hat{H}_B$ remain unchanged. As shown in Ref.~\cite{wojcik2025scaled}, the scaling approach accurately reproduces the lowest-energy two-electron states (singlet and unpolarized triplet), in which the electrons occupy different quantum dots, thereby minimizing the Coulomb repulsion.

In Fig.~\ref{fig:scaled}, we present the time evolution of the spin components during the SWAP operation for two representative QD sizes: (a, c, e) $d=11$~nm and (b, d, f) $d=33$~nm. For $d=11$~nm, the scaled model with $k=2$ reproduces the results of the original model with very good agreement. The same scaling factor also works well for the larger dot ($d=33$~nm), presented in Fig.~\ref{fig:scaled}. While increasing the scaling factor to $k=3$ leads to noticeable deviations from the original model, the agreement remains qualitatively good, particularly for the SWAP dynamics of $S_z$. These results indicate that quantum gates based on spin qubits realized in LAO/STO interfaces can be efficiently simulated using the scaled Hamiltonian, which significantly accelerates the time required for calculations.

\section{Summary}
\label{sec:summary}
In this work, we have presented a systematic theoretical study of spin dynamics and the SWAP operation in double QDs defined at the LaAlO$_3$/SrTiO$_3$ interface. By employing a real-space tight-binding model and the CI method, we have investigated how the multiorbital nature of the $t_{2g}$ conduction band and spin--orbit coupling influence the fidelity of spin exchange.

Our analysis revealed two distinct regimes governed by the size of the QDs:
(i) large quantum dots, for which the electronic structure is dominated by $d_{xy}$ orbitals and the spin dynamics are characterized by a Rashba-type SO interaction; in this regime, the SWAP operation is strongly perturbed by SO coupling; and (ii) small quantum dots, in which $d_{xz}$ and $d_{yz}$ orbitals provide a significant contribution to the lower-lying excited states; in this case, interorbital coupling induces irregular spin oscillations and beating patterns, significantly lowering the fidelity of the SWAP operation.

We identified an intermediate regime as a ``sweet spot'' where high-fidelity SWAP operations are possible. In this regime, the SO interaction is sufficiently low compared to the kinetic energy, and the orbital character remains relatively pure, allowing for nearly perfect spin exchange. Additionally, we explored the anisotropy of the SWAP operation, finding that fidelity can be optimized when the initial spin polarization is aligned with the effective SO field .

Finally, we validated a scaled tight-binding model, demonstrating it to be a computationally efficient and accurate tool for simulating multi-electron dynamics in transition-metal-oxide-based quantum gates.

\section*{Acknowledgments}
We gratefully acknowledge Roberta Citro for valuable discussions.
This work is financed by the Horizon Europe EIC Pathfinder under the grant IQARO number 101115190 titled "Spin-orbitronic quantum bits in reconfigurable 2D-oxides". We gratefully acknowledge Poland’s high-performance computing infrastructure PLGrid (HPC Center Cyfronet) for providing computer facilities and support within computational grant no. PLG/2025/018852.

\appendix
\section{Hamiltonian reduced to the $d_{xy}$ orbital}
\label{sec:a1}
By applying the standard folding-down transformation, the full Hamiltonian in Eq.~(\ref{eq:Hamiltonian_k_space}) can be mapped onto an effective Hamiltonian that acts solely within the subspace of the $d_{xy}$ electrons. The corresponding transformation is given by
\begin{equation}
    \hat{H}^{eff}_{xy} = \hat{H}_{xy} + \hat{H}_{c}(\hat{H}_{xz/yz} - E)^{-1}\hat{H}_{c}^\dagger.
    \label{Heff}
\end{equation}
where 
\begin{equation}
\hat{H}_{xy}=
\left(
\begin{array}{cc}
\epsilon^{xy}_{\mathbf{k}}  & 0\\
 0 & \epsilon^{xy}_{\mathbf{k}} 
\end{array} \right),
\end{equation}
and
\begin{equation}
\hat{H}_{xz/yz}=
\left(
\begin{array}{cccc}
\epsilon^{xz}_{\mathbf{k}} & 0 & 0 & 0 \\
 0 & \epsilon^{xz}_{\mathbf{k}} & 0 & 0 \\
 0 & 0 & \epsilon^{yz}_{\mathbf{k}} & 0 \\ 
 0 & 0 & 0 & \epsilon^{yz}_{\mathbf{k}} \\
\end{array} \right),
\end{equation}
where in the latter term, we neglect both the coupling of the \(d_{xz}/d_{yz}\) bands to the magnetic field and their hybridization, under the assumption that the kinetic and SO energy provide the dominant contributions to the total energy.
The coupling between the $d_{xy}$ and $d_{yz}/d_{xz}$ orbitals is given by
\begin{eqnarray}
\hat{H}_{c}&=&
\frac{\Delta_{SO}}{3}
\left(
\begin{array}{cccc}
0 & i & 0 & -1 \\
i & 0  & 1 & 0   
\end{array} \right)  \\
&+&i\Delta_{RSO}
\left(
\begin{array}{cccc}
\sin k_y & 0 & \sin k_x & 0 \\
0 & \sin k_y & 0 & \sin k_x   
\end{array} \right). \nonumber
\end{eqnarray}

Assuming that \(\Delta_E\) represents the largest energy scale in the system and taking the bottom of the \(d_{xy}\) band as the reference energy, we can expand $(\hat{H}_{xz/yz} - E)^{-1}$ from Eq.~(\ref{Heff}) in the Taylor series 
\begin{equation}
(\hat{H}_{xz/yz} - E)^{-1} = \frac{1}{\epsilon^{xz/yz}_{\mathbf{k}}+\Delta_E-E} \mathds{1}_{4\times4} \approx \frac{1}{\Delta_E}\mathds{1}_{4\times4}.
\end{equation}
Then, the effective Hamiltonian is reduced to the following form
\begin{eqnarray}
\hat{H}^{eff}_{xy}&=&\left ( \epsilon^{xz}_{\mathbf{k}} + \frac{2\Delta_{SO}\gamma}{3(1-\gamma)}\right ) \mathbf{1}_{2\times2}+\frac{1}{2}g\mu_B \mathbf{B}\cdot \pmb{\sigma} \nonumber \\ 
&+&\alpha (\sigma_y \sin k_x - \sigma _x \sin k_y)
\label{eq:Hamiltonian_k_xy}
\end{eqnarray}
where $\gamma=\Delta_{SO} / 3\Delta _E$ and $\alpha=\Delta_{SO}\Delta_{RSO}/3\Delta_E$. The last term in Eq.~(\ref{eq:Hamiltonian_k_xy}) is related to the SO coupling of the Rashba type \cite{Rashba2003} similar to that observed in semiconductors.

%\bibliography{bibliography}

\end{document}